\documentclass[useAMS,usenatbib]{mn2e}
\usepackage{graphicx}
\usepackage{amsmath,amssymb}

\def\ergscc{~{\rm erg ~ cm^{-2} ~ s^{-1}}}
\def\ergs{~{\rm erg ~ s^{-1}}}
\def\etal{{\it et~al.}}

\def\keV{~{\rm keV}}
\def\chandra{{\it Chandra}}
\def\cm2{~{\rm cm^{-2}}}
\begin{document}
\title[Modeling the Spectrum of IGR J17177-3656]{Modeling the Multi-band Spectrum of IGR J17177-3656}
\author[R. Ma]{Renyi Ma \thanks{E-mail: ryma@xmu.edu.cn} \\
Department of Physics and Institute of Theoretical Physics and Astrophysics, Xiamen University, Fujian 361005, China}

\date{Accepted 0000 . Received 0000; in original form 0000}

\pagerange{\pageref{firstpage}--\pageref{lastpage}} \pubyear{}

\maketitle

\label{firstpage}

\begin{abstract}
The correlation between radio and X-ray luminosity in the hard state of black hole X-ray binaries is important for unveiling  the relation between the accretion flow and the jets.
In this paper, we have modeled the quasi-simultaneous multi-band observations of a recently discovered transient X-ray source, IGR J17177-3656.
It is found that the source is probably an outlier following the steep radio/X-ray correlation rather than an outlier in the transition region as suggested by Paizis et al. (2011).
It is also found that the multi-band spectrum can be successfully
modeled by the luminous hot accretion flow (LHAF) but less likely by the
advection dominated accretion flow (ADAF).
Our results support the point that LHAF can explain the steep radio/X-ray correlation.
\end{abstract}


\section{INTRODUCTION}

The hard state of the black hole (BH) X-ray binaries is characterized by
a power-law X-ray spectrum with photon index $\Gamma=1.5-2.1$, and
relatively strong radio emission \citep[e.g.][]{ZdziarskiG04,RemillardM06, Fender06, DoneGK07}.
During the hard states, there exists an interesting
nonlinear correlation between the radio and X-ray luminosity,
i.e. $L_{radio}\propto L_X^b$ with $b \sim 0.5-0.7$
\citep[e.g.][]{Hannikainenetal98,Corbeletal00, Corbeletal03, Galloetal03}.
Although this relation is established from a few BH binaries like GX 339-4 and V404 Cyg,
it even holds for active galactic nuclei \citep{Merlonietal03, Falckeetal04, Gultekinetal09}.
Moreover, a similar correlation between optical/infrared and X-ray flux has been found \citep{Homanetal05,Russelletal06,Coriatetal09}.
It seems that this correlation is universal and can be taken as ``standard''.

However, radio observations of some new BH transients in more recent outbursts in the following years reveal that their behavior deviates significantly from the aforementioned correlation \citep[e.g.][]{Corbeletal04,
Rodriguezetal07, XueC07, Solerietal10}.
Based on the first precise measurement of an outlier H1743-322,
\citet[][{\bf C11}]{Coriatetal11} discovered a new tight correlation in its bright hard state, with the slope being $1.38\pm 0.03$.
At low luminosity H1743-322 rejoins the standard correlation with $b \sim 0.6$.
At the intermediate luminosity, there exists a transition region.
The reasons for the existence of the two correlation and their transition remain to be clarified.


To date three classes of models have been proposed to explain
the hard states: (i) the hot accretion flow, which includes ADAF \citep[e.g.][]{NY94, A95, C95,N05, Y07, NM08} and LHAF \citep[e.g.][]{Y01,Y03,Y07b,YZ04},
(ii) the disk corona \citep[e.g.][]{LiangP77,
HaardtM91, Zhang00, Liuetal02, Liuetal07, Liuetal11, ReisFM10},
and (iii) the jets  \citep{Markoffetal01, Markoffetal03, Reigetal03, Gianniosetal04,Maitraetal09,PeerC09}.
All these models can explain the ``standard'' correlation between radio and X-ray emission.
In the models of hot accretion flow and corona, X-rays come from the accretion flow while radio emission from the jet. Since the jet originates from the accretion flow, it is natural to expect the radio/X-ray correlation \citep[e.g.][]{Merlonietal03,YC05,YuanYH09}.
In the jet model, both radio and X-ray emission come from the same electrons in the jet, and thus strong correlation is expected \citep[e.g.][]{Markoffetal03,Falckeetal04}.
As for the  steep correlation with index of $\sim 1.4$, \citet{Coriatetal11} proposed {\bf that the efficient accretion flow like LHAF or disk corona can explain it}.
To understand the physics behind the correlation, more sample of sources are needed.

Very recently, \cite{P11} ({\bf P11}) reported the observation of a newly discovered X-ray transient, IGR J17177-3656.
They analyzed the quasi-simultaneous data by \chandra, INTEGRAL and ATCA, and argued that the source is probably a low-mass BH X-ray binary in the hard state, although the neutron star nature cannot be excluded yet. By assuming the distance to be 8~kpc, they found that the source follows the behavior of the BHC H1743-322 at intermediate X-ray luminosity.
In this paper, by using an accretion-jet model, we obtain the source distance by modeling the observed multi-band spectrum, and re-investigate its position in the radio/X-ray correlation diagram.
It is found that IGR J17177-3656 probably follows the steep radio/X-ray correlation of H1743-322 and not the transition one.

\section{MODELING THE SPECTRUM}
\subsection{Observational data}
After the discovery by INTEGRAL, IGR J17177-3656 became a focus of radio, NIR, and X-ray
telescopes ({\bf P11} and references therein).
\chandra ~ conducted an observation of 20 ks at March 22nd, 2011.
The 2-8keV observation shows a power-law spectrum, $\Gamma=1.36^{+0.16}_{-0.15}$,
with unabsorbed 3-9~keV flux being $7.5 \times 10^{-11} \ergscc$.
A simultaneous INTEGRAL observation in 20-200 keV shows
a power law with $\Gamma=1.8^{+0.9}_{-1.4}$
 and flux being $6.48^{+0.94}_{-1.15}\times 10^{-10} \ergscc $. This result is similar to the observation on March 15th, 2011.
The radio observation on the same day is conducted by Australia Telescope Compact Array (ACTA) at two frequency bands, which are centered at 5.5 GHz and 9 GHz.
The corresponding flux density are $0.24\pm0.06$ mJy and $0.20 \pm 0.06$ mJy, respectively. Figure~1 shows the results of our modeling and the data points of {\bf P11}.

\subsection{The accretion-jet model}

The accretion-jet model is briefly described here. The reader can refer to \citet{YCN05} for additional details.
There are four components in the model, the inner hot accretion flow (ADAF/LHAF), the outer standard accretion disk, the outflow and the jet.
In the ADAF regime, the accretion rate $\dot{M} \lesssim \alpha^2 \dot{M}_{Edd}$, where $\dot{M}_{Edd}=10L_{Edd}/c^2$ and $\alpha$ is the viscous parameter,  and the viscously dissipated energy can be advected inwards with the accreting gas.
While in the LHAF regime, $\alpha^2 \dot{M}_{Edd} \lesssim
\dot{M} \lesssim 5\alpha^2 \dot{M}_{Edd}$, and the radiative cooling is a little more efficient than the viscous heating.
Although the temperature of LHAF is lower than that of ADAF, it is also high because during the falling the compression work plus viscous heating together balance the radiative cooling.
The jet is shown in the flat/inverted radio spectra and the direct images
\citep[e.g.][]{Fender01,Fender06,Dhawanetal00,Stirlingetal01}.
The existence of outflow in ADAF/LHAF is from theoretical work or numerical simulations
\citep[e.g.][]{IgumenshchevA99,NY94, BB99,Netal00,YB10}.
The outer standard accretion disk is needed when modeling the UV spectra of some well-known BHs
\citep[e.g.][]{NMY96,Esinetal01, Fronteraetal01}.

The radiation processes in the hot accretion flow include synchrotron, bremsstrahlung, and inverse Compton scattering of soft photons.
The emission from the outer cool disk is modeled as a multi-color blackbody spectrum.
And in the jet, synchrotron emission and self-absorption of the relativistic electrons are considered.

The accretion disk is described by a set of parameters. The inner hot accretion flow truncates the outer standard accretion disk at the transition radius $r_{tr}$.
Since the physics of the transition is highly uncertain, the range of the value of $r_{tr}$ is large, from $\sim 10 r_S$ to $\sim 10^4 r_S$, where $r_S=2GM/c^2$ \citep[e.g.][]{EsinMN97}.
Due to the lack of UV spectra, we set it a normal value, $r_{tr}=100r_S$.
The influence of outflow on the accretion is mimicked with the parameter $s$ by $\dot{M}=\dot{M}_0 (r/r_{tr})^s$, where $\dot{M}_0$ is the accretion rate at $r_{tr}$.
The value of $s$ describes the strength of the outflow and varies in the range $0 < s < 0.5$.
For the case $s=0$, there is no outflow, while for the case $s=0.5$,
90\% of the accreted gas will flow away and only 10\% finally fall into the BH given $r_{tr}=100r_S$, which roughly agrees with numerical simulations \cite[e.g.][]{StoneP01}.
A recent numerical simulation shows that the fraction of outflow is almost the same for ADAF and LHAF \citep{YB10}. So in this paper, the value of $s$ is assumed to be the same for different accretion rates, instead of a function related to accretion rate as assumed in \citet{YCN05}.
To determine the dynamics of the hot accretion flow, we need three more parameters, which are associated with the MHD turbulence.
These parameters are the viscous parameter $\alpha$, the fraction of gas pressure to total pressure $\beta$, the fraction of the viscous dissipation directly heating electrons $\delta$.
The value of $\beta$ is fixed at a typical value of 0.9,
the value of $\alpha$ changes in the range $0.01-0.3$, and $\delta \sim 0.01-0.5$ \citep{QuataertG99}.



The jet is modeled phenomenologically with quite a few parameters.
We assume a fraction of the accretion flow, $\dot{m}_{jet}$, is transferred into the vertical direction, and the jet has a conical geometry with a small half-opening angle $\psi$.
For simplicity we assume each ejection moves at a constant velocity along the jet.
Different ejections may have different velocities, however, their averaged bulk Lorentz factor, $\Gamma_j$, is assumed to be constant.
When different ejections collide, internal shocks occur,
which accelerate a fraction of the electrons into power-law energy distribution with index $p$.
The energy density of accelerated electrons and amplified magnetic field are specified with
$\epsilon_e$ and $\epsilon_B$.
\citet{YC05} have shown that the contribution of the jets to the X-ray emission in BH sources is negligible when the luminosity is $ > 10^{-6}L_{Edd}$.
Therefore in the transient source IGR J17177-3656, the jet contributes only to the radio emission.
It should be noted that the main results of the paper are from the modeling of the X-ray observation, and we model the radio spectra just to show the ability of the model to explain the multi-band spectra. So we fix most of the jet parameters.
Considering that there are only two radio data points,
we leave $\dot{m}_{jet}$ and $p$ free and
the other parameters fixed at the value that has been used for other BH binaries, i.e. $\psi=0.1$, $\Gamma_j=1.2$, $\epsilon_e=0.06$ and $\epsilon_B=0.02$ \citep{YCN05,ZhangYC10}.


The BH mass is set it to be of normal value $M_{BH}=10 M_\odot$.
The binary inclination angle $\theta$ is estimated to be high $60^\circ < \theta < 75^\circ$ \citep{P11}.
Detailed calculations show the influence of $\theta$ is not significant, so
we set it to be a middle value $\theta = 70^\circ$.

There are 7 free parameters and quantities in the model,
$\dot{M}$, $s$, $\alpha$, $\delta$, $\dot{m}_{jet}$, $p$, and the distance $d$.
The modeling steps are as follows.
First, we try with different value of $\dot{M}$, $s$, $\alpha$, and $\delta$ to model the slopes of the X-ray spectra, and obtain the possible ranges of these parameters.
Second, the range of $d$ is obtained by comparing the observed X-ray flux and theoretically calculated disk luminosity.
At last, $\dot{m}_{jet}$ and $p$ are adjusted to check whether the radio spectrum can be modeled simultaneously.

\subsection{Modeling results}

\begin{figure}
\begin{center}
\includegraphics[width=65mm]{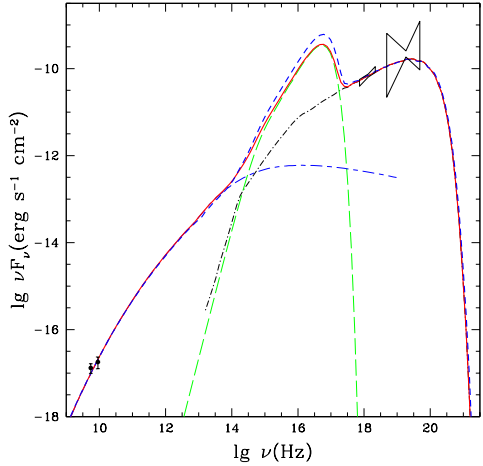}
\caption{Spectral modeling results for the quasi-simultaneous observation of
IGR J17177-3656 by ATCA (5GHz, 9GHz), \chandra (3-9\keV) and INTEGRAL (20-200\keV).
The {\it thick solid} line is the total spectrum and corresponds to Case I with $s=0.2$, $\dot{M}=0.4\dot{M}_{Edd}$, and $d=25$~kpc.
For Case I, the {\it short dash-long dashed}, {\it long dashed}, and {\it dot-dashed} lines correspond to radiation from the jet, the outer thin disk, and the inner hot accretion flow, respectively.
The {\it thick dashed} line is the total spectrum and corresponds to Case II with $s=0.3$, $\dot{M}=0.5\dot{M}_{Edd}$, and $d=21$~kpc.
}
\end{center}
\label{fsp}
\end{figure}

\begin{figure}
\begin{center}
\includegraphics[width=50mm]{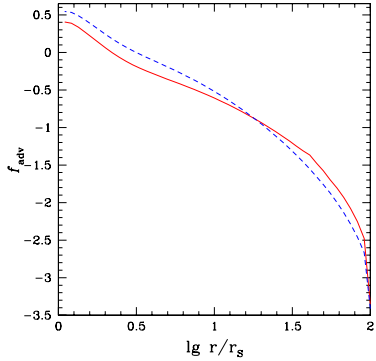}
\caption{The variation of the advection factor. The {\it solid} and {\it dashed} lines correspond to Case I and II, respectively.}
\end{center}
\label{fadv}
\end{figure}

\begin{figure}
\begin{center}
\includegraphics[width=65mm]{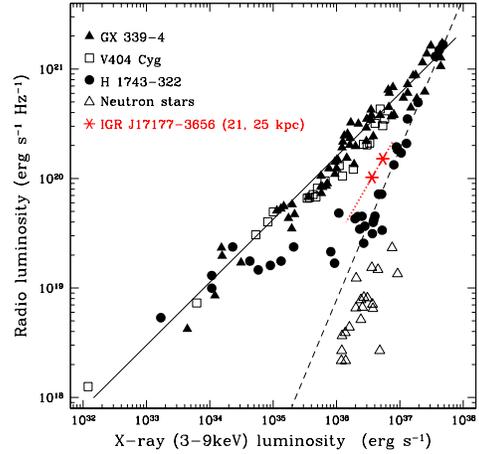}
\caption{Correlation between radio and X-ray luminosity for a sample of black holes and neutron stars. The asterisk and dotted line mark the possible location of IGR J17177-3656. {\bf The other data are adapted from {\bf C11}}. The solid and dashed lines are the fit with correlation index of 0.6 and 1.4.}
\end{center}
\label{fRX}
\end{figure}

It is useful to investigate the effects of parameters on the output spectra in order to
find the possible ranges of the parameters.
Note that the X-ray spectral index decreases with the increasing $y$ Compton parameter, which is the product of electron temperature and optical depth.
Given other parameters fixed, the surface density of the accretion flow of higher $\dot{M}$ is higher, therefore the optical depth is higher and the spectral index is smaller.
The value of $\alpha$ determines the maximum accretion rates of the ADAF and LHAF, and therefore the maximum optical depth and hardest spectra.
For larger $\delta$, more viscously dissipated energy is transferred to thermal electrons.
Consequently, the electron temperature is a little higher and the X-ray spectrum is harder.
When $s$ increases, the final X-ray spectrum is softer even if the electron temperature increases due to less cooling.
This is because, as shown by detailed calculations, the decrease of the surface density and optical depth are more significant than the increase of the electron temperature.
It should be mentioned that if the X-ray luminosity rather than $\dot{M}$ is given, the X-ray spectral hardness increases with $s$.
 This is because more photons come from the region of larger radius, where the optical depth is higher \citep{Y07b}.

Our detailed calculations show that in order to model the observed spectrum, the parameters are limited in the following ranges, $\dot{M} > 0.3 \dot{M}_{Edd}$,  $\alpha > 0.2$, $\delta > 0.1$, and $0.2 \lesssim s \lesssim 0.3$.
When $\dot{M} \lesssim 0.3 \dot{M}_{Edd}$, $\alpha \le 0.2$ or $\delta \lesssim 0.1$, the output photon index are greater than 1.54.
When $s < 0.2$, the obtained luminosity is as large as $8.1\times 10^{37} \ergs$, and the obtained distance is greater than 30~kpc, which seems unreasonable since the diameter of the Galaxy is about 30kpc.
When $s > 0.3$, the highest photon index obtained with $\dot{M}=5\alpha^2 \dot{M}_{Edd}$ is 1.65 , which is not hard enough to explain the observed X-ray spectra.

The modeling is not unique. As an example two cases of the modeled spectra are shown in Figure~1.
In Case I, the set of parameters are
$s=0.2$, $\alpha=0.3$, $\delta=0.5$, $\dot{M}=0.4\dot{M}_{Edd}$, $\dot{m}_{jet}=0.3\%$,  $p=2.25$, and $d=25$~kpc.
While in Case II, $s=0.3$, $\dot{M}=0.5\dot{M}_{Edd}$, and $d=21$~kpc, with others being the same as Case I.
The thick solid and dashed lines correspond to Case I and II, respectively.
From this figure it can be seen that the radio and X-ray emission are dominated by the jet and the hot accretion flow, respectively. The infrared and optical spectra are the sum of the emissions from the jet, hot accretion flow and the truncated thin disk.
It can also be seen that the model parameters can be further constrained by optical/UV observation.

The accretion rate is in the range $\alpha^2 \dot{M}_{Edd}< \dot{M} < 5\alpha^2 \dot{M}_{Edd}$,
therefore the accretion flow is in the regime of LHAF.
This regime is characterized by the negative advection factor, which is defined as $f_{adv}=q_{adv}/q^+=(q^+-q_{ie})/q^+$, with $q^+$, $q_{adv}$, $q_{ie}$ being the viscous heating rate, advection cooling rate, and the cooling rate of ions by ion-electron coupling, respectively. It is easy to find in Figure~2 that $f_{adv} < 0$ except in the innermost region $r < 6r_S$.
The key reason why LHAF can model the spectra but ADAF cannot is because of the observed hard X-ray spectra with {\bf photon index being  $\sim 1.36^{+0.16}_{-0.15}$}, which demands optically thick hot gas that is difficult for ADAF to form.


The distance is 21-25~kpc, rather large compared to most of the observed
BH X-ray binaries. However, this is still smaller than that of the farthest
dynamically confirmed BH GS 1354-64, which is estimated to lie at
25-60 kpc \citep{ReynoldsM11}. Therefore it is still well
acceptable. Considering the diameter of the Galaxy is 20-30 kpc, we think that the source may be located on the side of the Galaxy opposite to the sun.
Moreover, if the mass of the BH is adjustable, the distance could be as close as $\sim$13-15
kpc for $M_{BH}=3M_\odot$.



\section{DISCUSSION}
\subsection{Location of IGR J17177-3656 in radio/X-ray correlation diagram}


From the spectral modeling results, the distance of the source should be 13-30~kpc by considering all possible BH mass and the radius of the Galaxy.
The possible position of IGR J17177-3656 is shown in Figure~3.
It can be seen that except for low mass BH the possible location of the source is close to the branch with $b\sim 1.4$.
This indicates that IGR J17177-3656 is a new source like H1743-322, which follows the steep radio/X-ray luminosity correlation.
Moreover, our results show that LHAF can not only explain the relation between the radio and X-ray luminosity, but also model the multi-band spectrum.
This supports the result of {\bf C11} that LHAF is a candidate to explain the correlation index with $b\sim 1.4$.

The observation of H1743-322 shows that the correlation index is $\sim 0.6$ at low luminosity and
$\sim 1.4$ at high luminosity.
The change of the correlation index can be naturally explained by virtue of the accretion-jet model.
When $\dot{M}<\alpha^2 \dot{M}_{Edd}$, the hot accretion flow is in the regime of ADAF, the luminosity is low and the index is $\sim 0.6$ \citep{YC05}.
When $\dot{M}$ increases within a few times, the hot accretion flow is in the regime of LHAF, the luminosity is higher and the index is $\sim 1.4$.

LHAF is more luminous than ADAF, therefore it is used to explain luminous hard state with $L_X > 0.1 L_{Edd}$. A luminous hard state occurs during the rising part of an outburst, corresponding to the upper-right corner in the hardness-intensity diagrams (HIDs).
Considering the above results it is natural to expect prevalent steep radio/X-ray correlation in the luminous hard state.
Some observations seem to support this result (Yuan, private communication).


It should be noted that in addition to the change of correlation index at
high luminosity, the accretion-jet model predicts another change of the
correlation index at low luminosity. \citet{YC05} predicted that
when $L\la 10^{-6}L_{\rm Edd}$,  the X-ray emission should be dominated by
the jet rather than the accretion flow. Consequently the correlation index
changes from b$\sim 0.6$ to  $\sim 1.23$. This prediction was well confirmed by
later direct observations of AGNs \citep{WrobelTH08,Pellegrinietal07,YuanYH09} and was compatible with theoretical works based on observational data \citep{WuYC07,Pszotaetal08,YuYH11,Rodrigoetal12}.
\citet{Galloetal06}, however, found that the radio and X-ray data of the quiescent
state of A 0620-00 are consistent with an extrapolation of the V 404 Cyg and
GX 339-4 correlations. Since the exact track of A 0620-00 in outburst is still unknown \citep{Corbeletal08,YuanYH09}, it is doubtful to say this result conflicts with the prediction of \citet{YC05}.
\citet{Corbeletal08} reported the radio and X-ray observations to the quiescent state of
V404 Cyg and found that the correlation of this source remain the original
$b\sim 0.6$ one. This result may also not contradict with the 1.23 correlation, considering the quiescent luminosity of this source is very close to the critical luminosity of $10^{-6}L_{\rm Edd}$ \citep{YuanYH09}.

\subsection{Limitations of the model}

It should be noted that the results of this paper depend on the observed slope of the X-ray spectrum.
In the above calculation we take the upper limit of the X-ray photon index to be 1.52, which is of 68\% confidence.
If the upper limit is extended to 1.6, which is of 90-99\% confidence as shown in Figure 4 of {\bf P11}, the above results are less robust.
In this case the ADAF with weak outflow, $s \lesssim 0.1$, can marginally model the X-ray spectrum.
Moreover, the minimal possible distance of IGR J17177-3656 reduces to 18~kpc, which means the source locates around the point where the correlation begins to deviate from the $b \sim 1.4$ branch.

The physics of the transition from the outer standard accretion disk to the inner hot accretion flow remains unclear. Although different scenarios may explain the transition, such as evaporation, turbulent diffusive heat transport and secular instability in the cold disk
\citep[][and references therein]{Y01}, the position and gas temperature cannot be estimated self-consistently.
In this paper, the transition radius is simply assumed to be 100$r_S$ and the ion temperature to be 0.5-0.8 times of the Virial temperature.

If the variation of IGR J17177-3656 is similar to H1743-322, there will be a transition region between the radiatively inefficient and efficient branches. This means that a radio-quenching or X-ray-increasing process occurs during the transition from ADAF to LHAF.
However, a physical explanation for such processes has not been well understood.

In this paper, the outflow only affects the density of the accretion flow. However, its contribution to the loss of angular momentum and energy can produce difference by a factor of few \citep{XieY08}. Moreover, with the existence of ordered magnetic field, the loss of angular momentum can be even more significant \citep{Buetal09}. These effects of outflow on the critical accretion rates and output spectrum will be studied in our future work.

Our modeling is based on the argument of {\bf P11} that IGR J17177-3656 is regarded as a BH X-ray transient in hard state. However, the neutron star nature of the source cannot be excluded yet. More observations on the source are needed to test these results.

\section{CONCLUSION}

In this paper we model the quasi-simultaneous multi-band spectrum of a newly discovered hard X-ray transient, IGR J17177-3656, using an accretion-jet model.
The model is almost the same as that in \citet{Y07b}, but with a constant outflow parameter $s$ considering the recent numerical simulation \citep{YB10}.
The required accretion rate is probably in the regime of LHAF,
and the distance is about 21-25 kpc for $M_{BH}=10M_\odot$.
All possible locations of IGR J17177-3656 in the radio/X-ray luminosity diagram are shown in Figure 3. This source is close to the steep correlation branch with $b\sim 1.4$.
The point of \citet{Coriatetal11} is supported, that LHAF can explain the steep radio/X-ray correlation.

\section*{ACKNOWLEDGMENTS}
The author thanks Feng Yuan and Ding-Xiong Wang for valuable comments and discussions.
The author is grateful to the anonymous referees for the constructive and very useful comments.
This work is supported by the National Natural Science Foundation of China under grants 11143003 and 10833002, the Natural Science Foundation of Fujian Province of China (No.2011J01023), and in part by the National Basic Research Program of China under grant 2009CB824800.

\label{lastpage}


\begin{thebibliography}{99}

\bibitem[Abramowicz \etal (1995)]{A95}
Abramowicz, M. A. \etal~ 1995, ApJ, 438, L37


\bibitem[Blandford \& Begelman (1999)]{BB99}
Blandford, R. D., \& Begelman, M. C. 1999, MNRAS, 303, L1


\bibitem[Bu \etal (2009)]{Buetal09}
Bu, D.-F., Yuan, F., \& Xie, F.-G. 2009, MNRAS,392, 325



\bibitem[Chen \etal (1995)]{C95}
Chen, X. \etal~ 1995, ApJ, 443, L61

\bibitem[Corbel \etal (2000)]{Corbeletal00}
Corbel, S. \etal~ 2000, A\&A, 359, 251

\bibitem[Corbel \etal (2003)]{Corbeletal03}
Corbel, S. \etal~ 2003, A\&A, 400, 1007

\bibitem[Corbel \etal (2004)]{Corbeletal04}
Corbel, S. \etal~ 2004, ApJ, 617, 1272

\bibitem[Corbel \etal (2008)]{Corbeletal08}
Corbel, S., Koerding, E., \& Kaaret, P. 2008, MNRAS, 389, 1697

\bibitem[Coriat \etal (2009)]{Coriatetal09}
Coriat, M. \etal~ 2009, MNRAS, 400, 123

\bibitem[Coriat \etal (2011)]{Coriatetal11}
Coriat, M. \etal~ 2011, MNRAS, 414, 677 ({\bf C11})

\bibitem[Dhawan \etal (2000)]{Dhawanetal00}
Dhawan, V., Mirabel, I. F., Rodriguez, L. F. 2000, ApJ, 543, 373

\bibitem[Done \etal (2007)]{DoneGK07}
Done, C., Gierli\'{n}ski, M., \& Kubota, A. 2007, A\&AR, 15, 1

\bibitem[Esin \etal (1997)]{EsinMN97}
Esin, A. A., McClintock, J. E., \& Narayan, R. 1997, ApJ, 489, 865

\bibitem[Esin \etal (2001)]{Esinetal01}
Esin, A. A. \etal~ 2001, ApJ, 555, 483

\bibitem[Falcke \etal (2004)]{Falckeetal04}
Falcke, H., K\"{o}rding, E., \& Markoff, S. 2004, A\&A, 414, 895

\bibitem[Fender (2001)]{Fender01}
Fender, R.P. 2001, MNRAS, 322, 31

\bibitem[Fender (2006)]{Fender06}
Fender, R. P. 2006, in Lewin W. H. G., van der Klis M., eds. Cambridge Astrophys. Ser. Vol. 39, Compact Stellar X-ray Sources. Cambridge Univ. Press, Cambridge, p. 381

\bibitem[Frontera \etal (2001)]{Fronteraetal01}
Frontera, F. \etal~ 2001, ApJ, 561, 1006



\bibitem[Gallo \etal (2003)]{Galloetal03}
Gallo, E., Fender, R. P., \& Pooley, G. G. 2003, MNRAS, 344, 60

\bibitem[Gallo \etal (2006)]{Galloetal06}
Gallo, E. \etal~ 2006, MNRAS, 370, 1351

\bibitem[Giannios \etal (2004)]{Gianniosetal04}
Giannios, D., Kylafis, N. D., \& Psaltis, D. 2004, A\&A, 425, 163

\bibitem[G\"{u}ltekin \etal (2009)]{Gultekinetal09}
G\"{u}ltekin, K. \etal~ 2009, ApJ, 706, 404

\bibitem[Hannikainen \etal (1998)]{Hannikainenetal98}
Hannikainen, D. C. \etal~ 1998, A\&A, 337, 460

\bibitem[Haardt \& Maraschi (1991)]{HaardtM91}
Haardt, F., Maraschi, L. 1991, ApJ, 380, L51


\bibitem[Homan \etal (2005)]{Homanetal05}
Homan, J. \etal~ 2005, ApJ, 624, 295

\bibitem[Igumenshchev \& Abramowicz (1999)]{IgumenshchevA99}
Igumenshchev, I. V. \& Abramowicz M. A. 1999, MNRAS, 303, 309





\bibitem[Liang \& Price (1977)]{LiangP77}
Liang, E. P., \& Price, R. H. 1977, ApJ, 218, 247

\bibitem[Liu \etal (2002)]{Liuetal02}
Liu, B. F., Mineshige, S., \& Shibata, K. 2002, ApJ, 572, L173

\bibitem[Liu \etal (2007)]{Liuetal07}
Liu, B. F. \etal~ 2007, ApJ, 671, 695

\bibitem[Liu \etal (2011)]{Liuetal11}
Liu, B. F., Done, C., \& Taam, R. E. 2011, ApJ, 726, L10

\bibitem[Maitra \etal (2009)]{Maitraetal09}
Maitra, D. \etal~ 2009, MNRAS, 398, 1638


\bibitem[Markoff \etal (2001)]{Markoffetal01}
Markoff, S., Falcke, H., \& Fender, R. 2001, A\&A, 372, L25

\bibitem[Markoff \etal (2003)]{Markoffetal03}
Markoff, S \etal~ 2003, A\&A, 397, 645

\bibitem[Merloni \etal (2003)]{Merlonietal03}
Merloni, A., Heinz, S., \& di Matteo, T. 2003, MNRAS, 345, 1057

\bibitem[Narayan \& Yi (1994)]{NY94}
Narayan, R., \& Yi, I. 1994, ApJ, 428, L13

\bibitem[Narayan (2005)]{N05}
Narayan, R. 2005, Ap\&SS, 300, 177


\bibitem[Narayan \etal (2000)]{Netal00}
Narayan, R., Igumenshchev, I. V., \& Abramowicz, M. A. 2000, ApJ, 539, 798

\bibitem[Narayan \& McClintock (2008)]{NM08}
Narayan, R., \& McClintock, J. E. 2008, New Astron. Rev., 51,733

\bibitem[Narayan \etal (1996)]{NMY96}
Narayan, R., McClintock, J. E., \& Yi, I. 1996, ApJ, 457, 821

\bibitem[Paizis \etal (2011)]{P11}
Paizis, A. \etal~ 2011, ApJ, 738, 183 ({\bf P11})


\bibitem[Pe'er \& Casella (2009)]{PeerC09}
Pe'er, A., \& Casella, P. 2009, ApJ, 699, 1919


\bibitem[Pellegrini \etal (2007)]{Pellegrinietal07}
Pellegrini, S. \etal~ 2007, ApJ, 667, 749

\bibitem[Pszota \etal (2008)]{Pszotaetal08}
Pszota, G. \etal~ 2008, MNRAS, 389, 423

\bibitem[Quataert \& Gruzinov (1999)]{QuataertG99}
Quataert, E., \& Gruzinov, A. 1999, ApJ, 520, 248


\bibitem[Reig \etal (2003)]{Reigetal03}
Reig, P., Kylafis, N. D., \& Giannios, D. 2003, A\&A, 403, L15

\bibitem[Reis \etal (2010)]{ReisFM10}
Reis, R. C., Fabian, A. C., \& Miller, J. M. 2010, MNRAS, 402, 836

\bibitem[Remillard \& McClintock (2006)]{RemillardM06}
Remillard, R. A., \& McClintock, J. E. 2006, ARAA, 44, 49

\bibitem[Reynolds \& Miller (2011)]{ReynoldsM11}
Reynolds, M. T., \& Miller, J. M. 2011, ApJ, 734, L17

\bibitem[Rodriguez \etal (2007)]{Rodriguezetal07}
Rodriguez, J. \etal~ 2007, ApJ, 655, L97

\bibitem[Rodrigo \etal (2012)]{Rodrigoetal12}
Nemmen, R. S., Storchi-Bergmann, T., \& Eracleous, M. 2012, ApJ, in press (arXiv: 1112.4640)

\bibitem[Russell \etal (2006)]{Russelletal06}
Russell, D. M. \etal~ 2006, MNRAS, 371, 1334


\bibitem[Soleri \etal (2010)]{Solerietal10}
Soleri, P. \etal~ 2010, MNRAS, 406, 1471

\bibitem[Stirling \etal (2001)]{Stirlingetal01}
Stirling, A. M. \etal~ 2001, MNRAS, 327, 1273

\bibitem[Stone \& Pringle (2001)]{StoneP01}
Stone, J. M., \& Pringle, J. E. 2001, MNRAS, 322, 461


\bibitem[Wrobel \etal (2008)]{WrobelTH08}
Wrobel, J. M., Terashima, Y., \& Ho, L. C. 2008, ApJ, 675, 1041

\bibitem[Wu \etal (2007)]{WuYC07}
Wu, Q., Yuan, F., \& Cao, X. 2007, ApJ, 669, 96

\bibitem[Xie \& Yuan (2008)]{XieY08}
Xie, F.-G., \& Yuan, F. 2008, ApJ, 681, 499

\bibitem[Xue \& Cui (2007)]{XueC07}
Xue, Y. Q., \& Cui, W. 2007, A\&A, 466, 1053


\bibitem[Yu \etal (2011)]{YuYH11}
Yu, Z., Yuan, F., \& Ho, L. 2011, ApJ, 726, 87

\bibitem[Yuan (2001)]{Y01}
Yuan, F. 2001, MNRAS, 324, 119

\bibitem[Yuan (2003)]{Y03}
Yuan, F. 2003, ApJ, 594, L99

\bibitem[Yuan (2007)]{Y07}
Yuan, F. 2007, in Ho L. C., Wang J-M., eds, ASP Conf. Ser. Vol. 373,
The Central Engine of Active Galactic Nuclei. Astron. Soc. Pac., San Francisco, p.95

\bibitem[Yuan \& Bu (2010)]{YB10}
Yuan, F., \& Bu, D-F. 2010, MNRAS, 408, 1051

\bibitem[Yuan \& Cui (2005)]{YC05}
Yuan, F., \& Cui, W. 2005, ApJ, 629, 408

\bibitem[Yuan \etal (2005)]{YCN05}
Yuan, F., Cui, W., \& Narayan, R. 2005, ApJ, 620, 905

\bibitem[Yuan \etal (2009)]{YuanYH09}
Yuan, F., Yu, Z., \& Ho, L. C. 2009, ApJ, 703, 1034

\bibitem[Yuan \& Zdziarski (2004)]{YZ04}
Yuan, F., \& Zdziarski, A. A. 2004, MNRAS, 354, 953


\bibitem[Yuan \etal (2007)]{Y07b}
Yuan, F. \etal~ 2007, ApJ, 659, 541

\bibitem[Zdziarski \& Gierli\'{n}ski (2004)]{ZdziarskiG04}
Zdziarski, A. A., \& Gierli{\'{n}}ski, M. 2004, Prog. Theor. Phys. Suppl., 155, 99

\bibitem[Zhang \etal (2010)]{ZhangYC10}
Zhang, H., Yuan, F., \& Chaty, S. C. 2010, ApJ, 717, 929


\bibitem[Zhang \etal (2000)]{Zhang00}
Zhang, S. N. \etal~ 2000, Science, 287, 1239
\end{thebibliography}
\end{document}